\documentstyle[epsfig,12pt]{article}

\title{
{\bf Antimatter from Microscopic Black Holes}}
\author{{R. G. Daghigh$^{1}$ and J. I. Kapusta$^{2}$} \vspace*{0.1in}\\
$^1${\it Physics Department and Winnipeg Institute for Theoretical Physics}\\
 {\it University of Winnipeg, Winnipeg, Manitoba, Canada R3B 2E9}\\
$^2${\it School of Physics and Astronomy, University of Minnesota}\\
 {\it Minneapolis, Minnesota 55455, USA}}

\date{}

\begin{document}

\maketitle

\begin{abstract}
The spectrum of positrons and antiprotons produced by the outflow of high temperature matter surrounding microscopic black holes is calculated for energies between five GeV and the Planck energy.  The results may be applicable for the last few hours and minutes of a microscopic black hole's lifetime.
\end{abstract}

Hawking radiation from black holes \cite{Hawk} is of fundamental interest 
because it involves the interplay between quantum field theory and the strong 
field limit of general relativity.  It is of astrophysical interest if black 
holes exist with sufficiently small mass that they explode rather than accrete 
matter and radiation.  Since the Hawking temperature and black hole mass are 
related by $T_H = m_{\rm P}^2/8\pi M$, where $m_{\rm P} = G^{-1/2} = 1.22\times 
10^{19}$ GeV is the Planck mass with natural units $\hbar = c = k_{\rm B} = 1$, 
a present-day black hole will evaporate and eventually explode only if $T_H > 
2.7$ K (microwave background temperature).  This requires    
$M < 4.6\times 10^{22}$ kg which is approximately 1\% of the mass of the Earth.
More massive black holes are cooler and therefore will absorb more matter and
radiation than they emit, hence grow with time.  Taking into account emission
of gravitons, photons, and neutrinos a critical mass black hole today has a
Schwarszchild radius of 68 microns and a lifetime of $2\times10^{43}$ years.

Following earlier work by Heckler \cite{Heckler} and by Cline, Mostoslavsky and Servant \cite{cline}, one of us showed that a self-consistent description of the emitted quanta was that of an outgoing fluid just marginally kept in local equilibrium. It required the assumption of sufficient initial particle interaction, and viscosity played a crucial role \cite{me}.  That paper was followed by an extensive numerical analysis of the relativistic viscous fluid equations by us \cite{us1}.  We also calculated the spectrum of high energy gamma rays expected during the last days, hours, and minutes of the black hole's lifetime in that same paper, and later calculated the spectrum of high energy neutrinos as well \cite{us2}.  We suggested that the most promising route for discovery of such microscopic black holes is to search for point sources emitting gamma rays of ever-increasing energy until suddenly the source shuts off.

In this brief report we use our previous results to calculate the emission of high energy antimatter, specifically positrons and antiprotons, from black holes with Hawking temperatures greater than 100 GeV and corresponding masses less than 10$^8$ kg.  It is at these and higher temperatures that new physics will arise.  Such a study is especially important in the context of searches for anomalous cosmic sources of high energy antimatter by current and future detectors \cite{detect1}-\cite{detect4}.  Previous notable theoretical studies in this area have been carried out by MacGibbon and Webber \cite{nu4} and by Halzen, Zas, MacGibbon and Weeks \cite{nu5}, who calculated the instantaneous and time-integrated spectra of positrons and antiprotons arising from the decay of quark and gluon jets. 

We show that antiprotons remain in chemical equilibrium down to the decoupling or freezeout temperature, $T_f$, which is on the order of 140 MeV.  Their abundance relative to lighter particles such as pions and muons is quite small because of the Boltzmann factor $\exp(-m/T_f)$.  Of course protons will be produced in equal abundance as antiprotons, but the universe is dominated by matter and these would not be seen as anomalous.  Antineutrons will also be produced in equal abundance, but for the most part they will not travel astrophysically interesting distances before beta decaying.
   
The source of positrons in the viscous fluid picture is quite varied.  Photons, pions, electrons, and muons remain in local thermal equilibrium down to the freezeout temperature $T_f$, as shown in \cite{us1}.  Positrons therefore can be directly emitted at this temperature.  Positrons also come from decays
involving pions and muons.  The relevant processes are (i) a thermal $\pi^+$ decays into a $\mu^+$ and $\nu_{\mu}$, followed by the muon decay $\mu^+ \rightarrow e^+ \nu_e {\bar \nu}_{\mu}$, and (ii) a thermal muon decays in the same way.

An extensive study of the production and annihilation rates of spin-1/2 baryons and antibaryons was performed in \cite{Igor,Pasi} based on a form of the 
fluctuation-dissipation theorem.  That set of studies included all baryons in the octet \cite{Igor} (n, p, $\Lambda$, $\Sigma^+$, $\Sigma^0$, $\Sigma^-$, $\Xi^0$, $\Xi^-$) and in the decuplet \cite{Pasi} ($\Delta^{++}$, $\Delta^+$, $\Delta^0$, $\Delta^-$, $\Sigma^{*+}$, $\Sigma^{*0}$, $\Sigma^{*-}$, $\Xi^{*0}$, $\Xi^{*-}$, $\Omega$).  We have used those formulas to compute the chemical relaxation time $\tau_{\bar{p}}$ for antiprotons in the temperature range of 100 to 200 MeV.  The result is shown in Fig. 1 in the form of $\tau_{\bar{p}}^{-1}$ in units of MeV.  The inverse relaxation time, or relaxation rate, is a steeply rising function of temperature.  Also shown in the figure is the volume expansion rate, $\theta$, of the fluid emerging from the black hole as computed in \cite{us1}.  According to this figure, the relaxation rate for antiprotons drops below the expansion rate in the vicinity of $T_{\rm QCD} = 170$ MeV, the temperature at which the system makes a transition from quarks and gluons to hadrons.  This is very similar to that for pions \cite{us1}.  Although there is some small uncertainty in all these rates, and indeed in $T_{\rm QCD}$ as well, it does suggest that antiprotons are emitted in chemical equilibrium at about the same temperature as pions.  Therefore we use the same numerical value of $T_f$ for both.

Calculation of the directly emitted positrons and antiprotons parallels very closely the analysis in \cite{us1,us2}.  Fermions emitted from the decoupling surface have a Fermi distribution in the local rest frame of the fluid.  The phase space density is
\begin{equation}
f(E') = \frac{1}{{\rm e}^{E'/T_f} +1} \, .
\end{equation}
The energy appearing here is related to the energy as measured in the rest frame 
of the black hole $E$ and to the angle of emission $\theta$ relative to the radial vector
\begin{equation}
E' = \gamma_f (E-v_f p\cos\theta)
\end{equation}
where $v_f$ is the radial flow velocity at the decoupling radius and $\gamma_f = 1/\sqrt{1-v_f^2}$.  No particles will emerge if the angle is greater than $\pi/2$.  Therefore the instantaneous distribution is
\begin{displaymath}
\frac{d^2N_{e^+}^{\rm fluid}}{dE dt} = 
4\pi r_f^2 \left(\frac{p^2}{2\pi^2}\right)
\int_0^1 d(\cos\theta) \cos\theta f(E,\cos\theta) =
\frac{2r^2_fT_f}{\pi u_f} p 
\sum_{n=1}^{\infty}\frac{(-1)^{n+1}}{n}
\end{displaymath}
\begin{equation} 
\left\{ \left( 1 - \frac{T_f}{n u_fp} \right)
\exp[-n \gamma_f (E-v_fp)/T_f]
+ \frac{T_f}{n u_fp} \exp[-nE\gamma_f/T_f]
\right\} \, ,
\end{equation}
where $r_f$ is the radius of the decoupling suface and $u = v \gamma$.  Since we are interested in black holes with Hawking temperature much greater than $T_f$, resulting in $\gamma_f \gg 1$, and in relativistic particles of energy much greater than both their mass $m$ and $T_f$, this expression simplifies to
\begin{equation}
\frac{d^2N_{e^+}^{\rm fluid}}{dE dt} = 
\frac{2r_f^2 T_f E}{\pi \gamma_f}
\ln \left( 1 + {\rm e}^{-E/2\gamma_f T_f} 
{\rm e}^{-\gamma_f m^2/2E T_f}\right) \, .
\end{equation}
If $E \gg \gamma_f m^2/T_f$, it simplifies even further.
\begin{equation}
\frac{d^2N_{e^+}^{\rm fluid}}{dE dt} = 
\frac{2r_f^2 T_f E}{\pi \gamma_f}
\ln \left( 1 + {\rm e}^{-E/2\gamma_f T_f} \right) \, .
\end{equation}
The dependence of the decoupling radius and flow velocity on the black hole mass or Hawking temperature was studied extensively by us earlier \cite{us1,us2}; they are
\begin{equation}
r_f \approx \frac{0.892}{T_f} \sqrt{\frac{T_H}{T_f}} \;\;\;\;\;\;\;\;
\gamma_f \approx 0.224 \sqrt{\frac{T_H}{T_f}} \, . 
\end{equation}
Although the formulas above are labeled for positrons, exactly the same ones apply to antiprotons.

The spectrum can be integrated over time, starting when the Hawking temperature is $T_0$ and ending when the black hole has completely disappeared.  The formulas parallel that in \cite{us1,us2} and will not be repeated here.  The interesting limit is when the positron or antiproton energy $E \gg 2 \gamma_f(T_0) T_f$.  In this limit the spectrum is
\begin{equation}
\frac{dN_{e^+}^{\rm fluid}}{dE} \rightarrow
2.9 \times 10^{-2} \frac{m_P^2 T_f}{E^4}
\label{direct}
\end{equation}
This $E^{-4}$ spectrum is characteristic of the viscous fluid description of the microscopic black hole wind.

The instantaneous spectrum of positrons arising from the decay of muons in thermal equilibrium until the decoupling temperature $T_f$ can be computed by folding together the spectrum of muons together with the decay spectrum of positrons.  Since the positron mass is very small in comparison to the muon mass, the result is identical to the spectrum of neutrinos calculated in \cite{us2}.  The answer is
\begin{eqnarray}
\frac{d^2N_{e^+}^{{\rm dir} \; \mu}}{dEdt}&=&
\frac{2r_f^2 T_fE}{3\pi\gamma_f}
\sum_{n=1}^{\infty}\Biggl \{ -{\rm Ei}
 \left( -\frac{nE}{2\gamma_fT_f}\right)\left[9\frac{nE}{2\gamma_fT_f}
+2\left(\frac{nE}{2\gamma_fT_f}\right)^2\right] \nonumber\\
&&+\exp\left(-\frac{nE}{2\gamma_fT_f}\right)\left[\frac{10\gamma_fT_f}{nE}-7-
2\frac{nE}{\gamma_fT_f}\right]
\Biggl \},
\end{eqnarray}
where ${\rm Ei}$ is the exponential-integral function.  In the high energy 
limit, defined here by $E \gg \gamma_fT_f$, the spectrum simplifies to
\begin{equation}
\frac{d^2N_{e^+}^{{\rm dir} \; \mu}}{dEdt}=\frac{2r_f^2T_fE}{3\pi\gamma_f}
\exp{\left(-\frac{E}{2\gamma_fT_f}\right)} \, .
\end{equation}
The time-integrated spectrum for neutrinos was calculated in \cite{us2}, and can be taken over directly for positrons.  It is a lengthy and not very interesting expression.  What is interesting is the high energy limit, where the spectrum simplifies to
\begin{eqnarray}
\frac{dN_{e^+}^{{\rm dir} \; \mu}}{dE} \rightarrow 
4 \times 10^{-3} \frac{m_P^2T_f}{E^4} \, .
\end{eqnarray} 

The spectrum of positrons coming from the decay of muons which themselves came from the decay of pions which were emitted at the decoupling temperature can also be taken over directly from \cite{us2}.  The instantaneous spectrum is
\begin{eqnarray}
\frac{d^2N_{e^+}^{{\rm indir} \; \mu}}{dEdt} &=&
\frac{m_{\pi}r_f^2 T_f^2}{3\pi q}
\sum_{n=1}^{\infty} \frac{1}{n^2} \Biggl\{
{\rm Ei}(-nx)
\left(5-\frac{27}{6}n^2x^2-\frac{2}{3}n^3x^3\right)
\nonumber\\
&-&\left(-\frac{19}{6}+\frac{23}{6}nx+\frac{2}{3}n^2x^2\right)
{\rm e}^{-nx}
\Biggl\}_{x_-}^{x_+}
\end{eqnarray}
where
\begin{eqnarray}
x_{\pm}=\frac{m_{\pi}E}{2\gamma_fT_f} \left[ 
\frac{(m_{\mu}^2+q^2)^{1/2}\pm q}{m_{\mu}^2}
\right]
\end{eqnarray}
and
\begin{equation}
m_{\pi} = q + \sqrt{m_{\mu}^2 + q^2} \, .
\end{equation}
The high energy limit $E \gg \gamma_fT_f$ is
\begin{eqnarray}
\frac{d^2N_{e^+}^{{\rm indir} \; \mu}}{dEdt} \rightarrow
\frac{m_{\pi}r_f^2T_f^2}{3\pi q}
\left[\frac{{\rm e}^{-x_-}}{x_-^2}-\frac{{\rm e}^{-x_+}}{x_+^2}\right] \, .
\end{eqnarray}
The integral over time cannot be expressed in closed form, is lengthy, and is not illuminating.  The high energy limit is simple and of the familiar form 
$1/E^4$.  
\begin{equation}
\frac{dN_{e^+}^{{\rm indir} \; \mu}}{dE} \rightarrow 
 8 \times 10^{-4} \frac{m_P^2T_f}{E^4}
\end{equation}

The instantaneous spectra are displayed in Fig. 2 for a Hawking temperature of 100 GeV corresponding to a black hole mass of $10^8$ kg and a lifetime of 5.4 days.  The spectrum of directly emitted electrons, positrons, protons, and antiprotons are identical above an energy of 5 to 10 GeV where the particle masses can be neglected.  These are the dominant source of antiparticles.  The next largest source of positrons is the decay of directly emitted muons.  This curve is lower because the energy of the muon is shared between the decay positron and the decay neutrino.  The smallest source of positrons is the decay of muons which themselves were the decay product of a pion.  This curve is the lowest because the energy of the directly produced pion is first split between the decay muon and the decay neutrino, and the energy is further degraded by sharing among the three decay products of the muon.  The overall shape of all three curves is exponential with the effective temperature $2 \gamma_f T_f \approx 0.24 \sqrt{T_H T_f}$, which in this example is about 1.7 GeV.

The time-integrated spectra, starting at the moment when the Hawking temperature is 100 GeV, are shown in Fig. 3.  The relative magnitudes and average energies reflect the trends seen in Fig. 2 and for the same reasons.  At high energy the spectrum is proportional to $E^{-4}$ for all three sources of antiparticles.  To get the total number of positrons one must add all three curves, while the top curve alone gives the yield of antiprotons.

We now turn to the possibility of observing antiparticles from a microscopic black hole directly.  Obviously this depends on many factors, such as the 
distance to the black hole, the size of the detector, the efficiency as a function of energy, how long the detector looks at the black hole before it is gone, how much matter lies along the line of sight to the black hole that could absorb or degrade the antiparticle beam, the strength of the magnetic field between the black hole and the detector that would deflect the charged particles, and so on.

For the sake of discussion, let us assume that one is interested in antiprotons 
with energy greater than a minimum of about 10 GeV and that the observational time is the last 5.4 days of the black hole's existence when its Hawking temperature is 100 GeV and above.  Integration of eq. (\ref{direct}) from $E_{\rm min}$ to infinity, and multiplying by 2 to account both for antiprotons emitted directly and for antiprotons that come from the eventual decay of antineutrons long after their emission from the black hole, results in the total number of
\begin{equation}
N_{\bar{p}} = \frac{m_P^2 T_f}{50 E_{\rm min}^3}
\approx 4\times 10^{32} \left( \frac{T_f}{140\;{\rm MeV}} \right)
\left( \frac{10\;{\rm GeV}}{E_{\rm min}} \right)^3 \, .
\end{equation}
For an exploding black hole located a distance $d$ from Earth the number of antiprotons in a detector with area $A$ is
\begin{equation}
N_{\bar{p}}(E > E_{\rm min}) = 
0.033 \left(\frac{1\;{\rm pc}}{d}\right)^2 
\left(\frac{A}{1\;{{\rm m}}^2}\right)
\left( \frac{10\;{\rm GeV}}{E_{\rm min}} \right)^3
\left( \frac{T_f}{140\;{\rm MeV}} \right) \, .
\end{equation}
A typical detector flown on a balloon or in Earth orbit would have an area of 1 m$^2$ and a viewing solid angle of 1 steradian.  This small number illustrates the difficulty of observing exploding black holes directly via antiprotons. The number of positrons would be approximately half this amount.  

What is the local rate density $\dot{\rho}_{\rm local}$ of exploding black 
holes?  This is, of course, unknown since no one has ever knowingly 
observed a black hole explosion.  The first observational limit was determined 
by Page and Hawking \cite{PH}.  They found that the local rate density
is less than 1 to 10 per cubic parsec per year on the 
basis of diffuse gamma rays with energies on the order of 100 MeV.  This limit 
has not been lowered very much during the intervening twenty-five years.  For 
example, Wright \cite{W} used EGRET data to search for an anisotropic
high-lattitude component of diffuse gamma rays in the energy range from 30 MeV 
to 100 GeV as a signal for steady emission of microscopic black holes.  He 
concluded that $\dot{\rho}_{\rm local}$ is less than about 0.4 per cubic parsec 
per year.  The number of antiprotons that could be created per cubic kilometer per billion years is
\begin{equation}
\dot{n}_{\bar{p}}(E > E_{\rm min}) = 14
\left(\frac{\dot{\rho}_{\rm local}}{1\;{{\rm pc}}^{-3}\;{\rm yr}^{-1}}\right)
\left( \frac{10\;{\rm GeV}}{E_{\rm min}} \right)^3
\left( \frac{T_f}{140\;{\rm MeV}} \right) \;
{\rm km}^{-3}\;({\rm billion \; yr})^{-1}\, .
\end{equation}
This is a very small number.  It would be unlikely that we could infer the existence of exploding microscopic black holes on the basis of diffuse antimatter. 

This paper has been a continuation of our previous work on a viscous fluid 
description of the radiation from microscopic black holes.  Here we 
have calculated the spectra of high energy positrons and antiprotons that would be emitted during the last few days of a black hole lifetime.  The flux of high energy antimatter would be small for black holes that explode at distances of the order of 1 parsec.  The spectrum does extend up to the Planck energy and could in principle be observed in high energy high altitude air showers.  However, unless the black hole was relatively close, one would not know whether the primary particle came from a black hole or from somewhere else.  High energy gamma rays from black holes \cite{us1} are probably easier to see and high energy neutrinos \cite{us2} are probably more interesting.  Nevertheless, if we were so fortunate to have a microscopic black hole explode not far away from the solar system the detection of all these types of particles might be a window to new physics beyond the 1 TeV scale.

\section*{Acknowledgements}

We are grateful to M. A. DuVernois for discussions about positrons and antiprotons in the high energy cosmic rays.  This work was supported by the US Department of Energy under grant DE-FG02-87ER40328 and by the Natural Science and Engineering Research Council of Canada.

\newpage

\begin{figure}
\centerline{\epsfig{figure=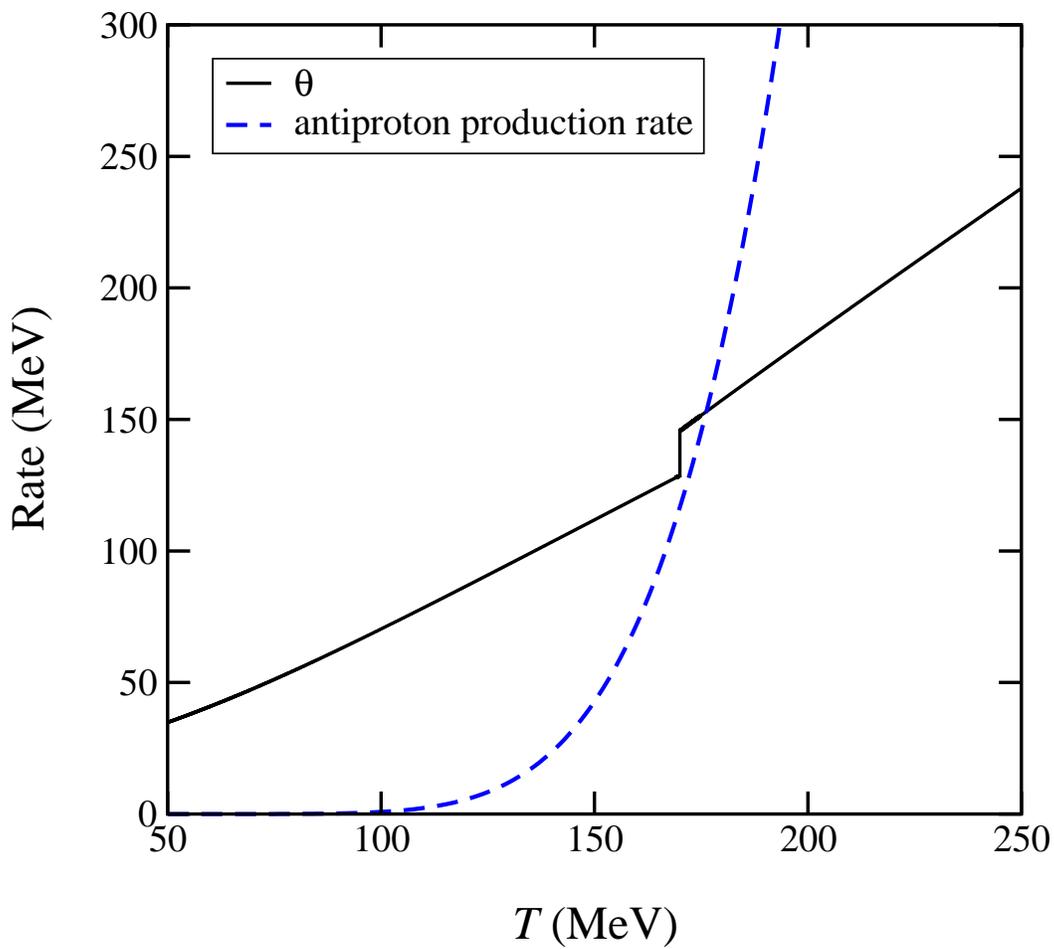,width=15.0cm,angle=270}}
\caption{The chemical equilibration rate for protons or anti-protons is compared to the local volume expansion rate.  The Hawking temperature is 10 TeV.}
\end{figure}

\begin{figure}
\centerline{\epsfig{figure=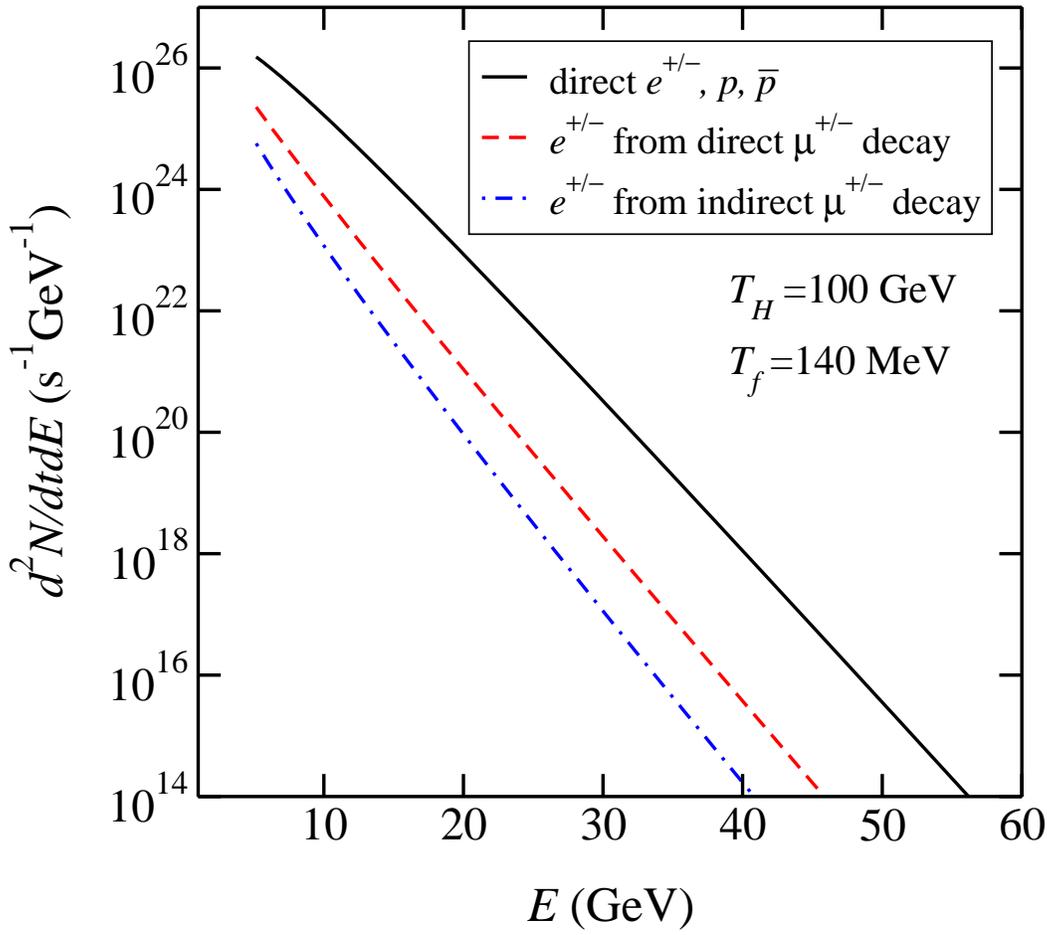,width=15cm,angle=270}}
\caption{The instantaneous direct electron, positron, proton and anti-proton spectra emerging from the fluid with decoupling temperature of $T_f=140$ MeV.  Also shown are electrons and positrons arising from direct muon and indirect muon decays.  Here the black hole temperature is $T_H=100$ GeV.}
\end{figure}

\begin{figure}
\centerline{\epsfig{figure=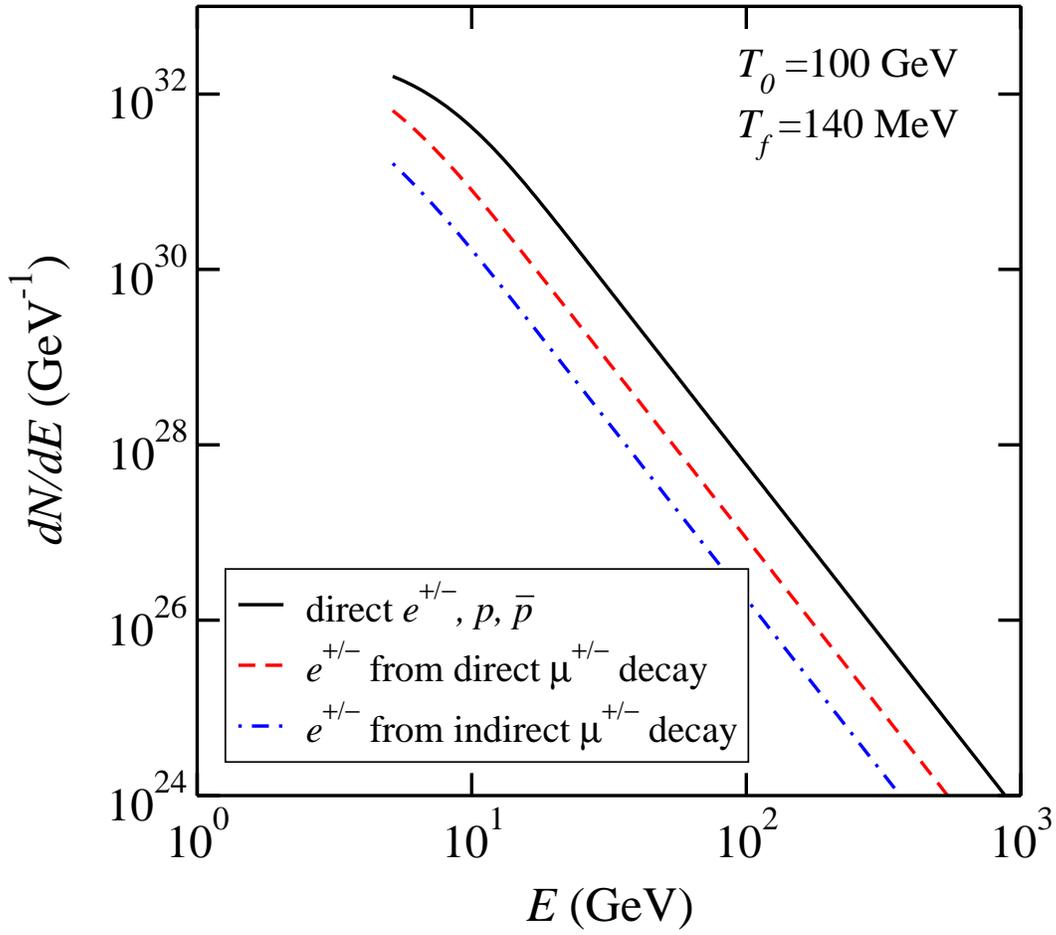,width=15cm,angle=270}}
\caption{The time integrated electron, positron, proton and anti-proton spectra emerging from a microscopic black hole.  Here the calculation begins when the black hole temperature is $T_0=100$ GeV.}
\end{figure}

\end{document}